\documentclass[12pt]{article}
\usepackage{amsmath,amsthm,amssymb,cite}
\newtheorem{thm}{Theorem}[section]
\newtheorem{prop}[thm]{Proposition}

\newtheorem{dfn}[thm]{Definition}

\newcommand{\nc}{\newcommand}
\nc{\mref}[1]{(\ref{#1})}
\nc{\vt}{v_{2\gL_0}}
\nc{\vo}{v_{\gL_0}}
\nc{\vot}{v_{\gL_1+\gL_0}}
\nc{\vw}{v_{\gL_1}}
\nc{\ppmm}{\genfrac{}{}{-10pt}{10pt}{++}{--}}
\nc{\wom}[5]{\Omega\left(\left.\begin{array}{ll}{#1}&{#2}\\{#3}&{#4}\end{array}\right|{#5}\right)}
\nc{\com}[5]{\chi\left(\left.\begin{array}{ll}{#1}&{#2}\\{#3}&{#4}\end{array}\right|{#5}\right)}
\nc{\we}[5]{W\left(\left.\begin{array}{ll}{#1}&{#2}\\{#3}&{#4}\end{array}\right|{#5}\right)}
\nc{\web}[5]{\overline{W}\left(\left.\begin{array}{ll}{#1}&{#2}\\{#3}&{#4}\end{array}\right|{#5}\right)}
\nc{\wep}[5]{W'\left(\left.\begin{array}{ll}{#1}&{#2}\\{#3}&{#4}\end{array}\right|{#5}\right)}
\nc{\wes}[5]{W^*\left(\left.\begin{array}{ll}{#1}&{#2}\\{#3}&{#4}\end{array}\right|{#5}\right)}
\nc{\wess}[5]{W^{**}\left(\left.\begin{array}{ll}{#1}&{#2}\\{#3}&{#4}\end{array}\right|{#5}\right)}
\nc{\cet}[7]{C^{#6}_{#7}\left(\left.\begin{array}{ll}{#1}&{#2}\\{#3}&{#4}\end{array}\right|{#5}\right)}
\nc{\bcet}[7]{\bar{C}^{#6}_{#7}\left(\left.\begin{array}{ll}{#1}&{#2}\\{#3}&{#4}\end{array}\right|{#5}\right)}
\nc{\wet}[7]{W^{#6}_{#7}\left(\left.\begin{array}{ll}{#1}&{#2}\\{#3}&{#4}\end{array}\right|{#5}\right)}
\nc{\bwet}[7]{\overline{W}^{#6}_{#7}\left(\left.\begin{array}{ll}{#1}&{#2}\\{#3}&{#4}\end{array}\right|{#5}\right)}
\nc{\wec}[7]{\widetilde{W}^{#6}_{#7}\left(\left.\begin{array}{ll}{#1}&{#2}\\{#3}&{#4}\end{array}\right|{#5}\right)}
\nc{\wgen}[6]{W^{#6}\left(\left.\begin{array}{ll}{#1}&{#2}\\{#3}&{#4}\end{array}\right|{#5}\right)}
\nc{\wgenp}[6]{W^{*{#6}}\left(\left.\begin{array}{ll}{#1}&{#2}\\{#3}&{#4}\end{array}\right|{#5}\right)}
\nc{\wo}[5]{\Omega\left(\left.\begin{array}{ll}{#1}&{#2}\\{#3}&{#4}\end{array}\right|{#5}\right)}
\nc{\wsgen}[8]{{#8}^{#6}_{#7}\left(\left.\begin{array}{ll}{#1}&{#2}\\{#3}&{#4}\end{array}\right|{#5}\right)}
\nc{\qbinom}[2]{{\genfrac{[}{]}{0pt}{}{{#1}}{{#2}}}_{q}}
\nc{\hg}[4]{{}_2\phi_1\left({{{#1}\,\,\,{#2}}\atop{{#3}}};p,
                     {#4}\right)}
\nc{\hhg}[4]{\phi\left({{{#1}\,\,\,{#2}}\atop{{#3}}};
                     {#4}\right)}
\nc{\fullhhg}[5]{{_1}\phi_2\left({{{#1}\,\,\,{#2}}\atop{{#3}}};
                     {#4},{#5}\right)}
\nc{\bra}[1]{\langle #1 |}
\nc{\ket}[1]{| #1 \rangle}
\nc{\qp}[2]{({#1}\, ; \, {#2})_{\infty}}
\nc{\qpf}[1]{({#1}\, ; \, q^4)_{\infty}}
\nc{\pp}[1]{({#1}\, ; \, p)_{\infty}}
\nc{\qpp}[1]{({#1}\, ; \, p, q^4)_{\infty}}
\nc{\sect}{\section}
\nc{\ssect}{\subsection}
\nc{\sssect}{\subsubsection}
\nc{\ud}[1]{\underline{{#1}}}
\nc{\isomo}{\buildrel {\sim} \over \longrightarrow}
\nc{\Aff}{\operatorname{Aff}}
\nc{\ot}{\otimes}
\nc{\br}[1]{\begin{array}{#1}}
\nc{\er}{\end{array}}
\nc{\bev}[1]{\begin{equation}\begin{array}{#1}}
\nc{\eeq}{\end{equation}}
\nc{\be}{\begin{eqnarray}}
\nc{\ee}{\end{eqnarray}}
\nc{\ben}{\begin{eqnarray*}}
\nc{\een}{\end{eqnarray*}}
\nc{\bec}{\begin{equation}\begin{array}{lll}}
\nc{\eec}{\end{array}\end{equation}}
\nc{\ed}{\end{document}}
\nc{\half}{\ensuremath{\frac{1}{2}}}
\nc{\Hom}{\operatorname{Hom}}
\nc{\End}{\operatorname{End}}
\nc{\vac}{|\textrm{vac}\rangle}
\nc{\dvac}{\langle\textrm{vac}|}
\nc{\id}{1}
\nc{\ra}{\rightarrow}  
\nc{\lra}{\longrightarrow}
\nc{\uqp}{U^{\prime}_q (\widehat{sl}_2)}
\nc{\uqbp}{U_q (b_+)}
\nc{\uqbm}{U_q (b_-)}
\nc{\ub}{U^{\prime}_q (b_+)}
\nc{\vsl}{V(\sigma(\lambda))}
\nc{\vl}{V(\lambda)}  
\nc{\bu}{\bullet}
\nc{\an}{{\ell}}
\nc{\slth}{\widehat{\mathfrak{sl}}_2\hskip 1pt}
\nc{\uq}{U_q(\slth)}
\nc{\ws}{\;\;}
\nc{\qu}{{1\ov 4}}
\nc{\hif}{\hb{ if }}
\nc{\hev}{\hb{ is even }}
\nc{\hod}{\hb{ is odd }}
\nc{\Tr}{{\rm Tr}}
\nc{\ad}{{\rm Ad}}
\nc{\hb}{\hbox}
\nc{\nn}{\nonumber} 
\nc{\curlra}{\buildrel{\sim}\over\longrightarrow}
\nc{\epp}{\varepsilon^{\prime}} 
\nc{\ol}{\overline}
\nc{\pl}{\prod\limits} 
\nc{\sli}{\sum\limits} 
\nc{\nin}{\noindent}

\nc{\ga}{\alpha}
\nc{\gb}{\beta}
\nc{\gd}{\delta}
\nc{\gep}{\varepsilon}
\nc{\gz}{\zeta}
\nc{\gt}{\theta}
\nc{\gk}{\kappa}
\nc{\gl}{\lambda}
\nc{\gp}{\phi}
\nc{\gs}{\sigma}
\nc{\go}{\omega}
\nc{\gn}{\nu}
\nc{\gr}{\rho}

\nc{\s}{\sigma}
\nc{\ep}{\varepsilon}
\nc{\z}{\zeta}
\nc{\g}{\gamma}
\nc{\zi}{\zeta^{-1}}

\nc{\gG}{\Gamma}
\nc{\gD}{\Delta}
\nc{\gT}{\Theta}
\nc{\gL}{\Lambda}
\nc{\gO}{\Omega}
\nc{\gP}{\Phi}

\nc{\cL}{\mathcal{L}}
\nc{\cF}{\mathcal{F}}
\nc{\cA}{\mathcal{A}}
\nc{\cP}{\mathcal{P}}
\nc{\cR}{\mathcal{R}}
\nc{\cS}{\mathcal{S}}
\nc{\cN}{\mathcal{N}}
\nc{\cD}{\mathcal{D}}
\nc{\cH}{\mathcal{H}}
\nc{\cO}{\mathcal{O}}
\nc{\cT}{\mathcal{T}}
\nc{\cQ}{\mathcal{Q}}
\nc{\cW}{\mathcal{W}}

\nc{\C}{\mathbb{C}}
\nc{\Q}{\mathbb{Q}}
\nc{\R}{\mathbb{R}}
\nc{\Z}{\mathbb{Z}}
\nc{\N}{\mathbb{N}}

\nc{\fg}{\mathfrak{g}}

\nc{\bi}{\bar{i}}
\nc{\bj}{\bar{j}}
\nc{\bp}{\bar{p}}
\nc{\bR}{\overline{R}}
\nc{\bgr}{\bar{\rho}}
\nc{\bA}{\bar{\alpha}}
\nc{\bB}{\bar{\beta}}
\nc{\bC}{\bar{\gamma}}
\nc{\by}{\bar{y}}

\nc{\tf}{\tilde{f}}
\nc{\te}{\tilde{e}}
\nc{\ts}{\tilde{s}}
\nc{\tgP}{\widetilde{\Phi}}
\nc{\tgPs}{\tilde{\Psi}}
\nc{\tgn}{\tilde{\nu}}
\nc{\tgl}{\tilde{\lambda}}
\nc{\tge}{\tilde{\eta}}
\nc{\txi}{\tilde{\xi}}

\nc{\cC}{\check{c}}
\nc{\cB}{\check{b}}

\nc{\goto}{\mapsto}
\nc{\embed}{\hookrightarrow}
\nc{\rien}{\emptyset}
\nc{\lb}[1]{\label{#1}}
\nc{\Nt}{\frac{N}{2}}
\nc{\vn}{\hspace*{-33truemm}}
\nc{\vm}{\hspace*{-0truemm}}
\nc{\ti}{t^{-1}}
\nc{\vb}{v^{(1)}}
\nc{\vbn}{v^{(n)}}
\nc{\us}{\underline{s}}
\nc{\vz}{V^{(1)}_z}
\nc{\vzn}{V^{(n)}_z}
\nc{\vzo}{V^{(1)}_1}
\nc{\piz}{\pi_z^{(1)}}
\nc{\pizn}{\pi_z^{(n)}}
\nc{\pis}{\pi_{(z,\us)}}
\nc{\bW}{\overline{W}}
\nc{\bQ}{\overline{Q}}
\nc{\tQ}{\widetilde{Q}}
\nc{\bT}{\overline{T}}
\begin{document}
\begin{flushright}
EMPG-02-13\\
\end{flushright}
\begin{center}
{\LARGE \bf A Generalized $Q$-operator for \\ $U_q(\widehat{sl_2})$
   Vertex Models\\[10mm] }
{\large \bf Marco Rossi\footnote{\tt M.Rossi@ma.hw.ac.uk} and Robert
   Weston\footnote{\tt R.A.Weston@ma.hw.ac.uk}}\\[6mm]
{\it  Department of Mathematics, Heriot-Watt University,\\
Edinburgh EH14 4AS, UK.}\\[5mm]
June 2002 - revised Sept 2002\\[10mm]
\end{center}
\begin{abstract}
\noindent 
In this paper, we construct a $Q$-operator as a trace of a 
representation of the universal 
$R$-matrix of $\uq$ over an infinite-dimensional auxiliary space. 
This auxiliary space is a four-parameter generalization
of the q-oscillator representations used previously. We
derive generalized $T$-$Q$ relations in which 3 of these parameters shift.
After a suitable restriction
of parameters, we give an explicit expression for the $Q$-operator of
the
6-vertex model and show the connection with Baxter's expression
for the central block of his corresponding operator. 
\end{abstract}
\nopagebreak

\section{Introduction}

Baxter's $Q$-operator has an interesting history. It was 
first constructed in 1972 as a tool for solving the 8-vertex model.
The background was that the 6-vertex model
had been solved by Bethe ansatz in the mid 60s by Lieb and Sutherland \cite{Lie67a,Lie67b,Lie67c,Sut67}.
However, this technique couldn't be simply extended
to the 8-vertex model due to the absence of a suitable Bethe ansatz
pseudo-vacuum (a problem associated with the lack of `charge
conservation' through vertices for this model). Then, in a seminal
series of papers 
\cite{Bax72a,Bax73aI,Bax73aII,Bax73aIII}, Baxter introduced his $Q$-operator\,\footnote{In fact, Baxter gives two different
constructions of a Q-operator in \cite{Bax72a} and
\cite{Bax73aI,Bax73aII,Bax73aIII}. We shall continue for the moment to use the generic term
`Baxter's $Q$-operator', and will specify to which construction we are referring
when
it becomes necessary to do so.}
as an apparent {\it deus ex machina} which allowed him to
write down Bethe equations for the eigenvalues of the 8-vertex model
transfer matrix {\it without} having an ansatz for the eigenvectors
(he did of course construct some eigenvectors using other techniques - see 
\cite{Bax73aI,Bax73aII,Bax73aIII}).

His approach was to start with the 6-vertex model Bethe ansatz, and to
derive certain functional relations between the transfer matrix $T(v)$
and a matrix $Q(v)$ - the elements
of both matrices being entire functions. He went on to show that the reverse
argument
could be used in order to start from the functional relations 
(and some other properties of $T(v)$ and $Q(v)$) and
derive the Bethe equations. He then considered the 8-vertex model,
constructed a $Q(v)$ operator that obeyed the correct requirements,
and used the reverse argument to derive Bethe equations. The approach
is described clearly in Baxter's book \cite{Bax82}.

Later on in the 70s, the quantum 
inverse scattering method (QISM) was developed and used to produce a rather
simpler derivation of the same Bethe equations for the 8-vertex model 
(the algebraic Bethe ansatz approach) \cite{tafa79}.
Baxter also invented his corner transfer matrix technique for the 8-vertex
model \cite{Bax82}. So, remarkably successful though it was, the $Q$-operator approach
perhaps came to be considered by many as a historical curiosity.

However, in the last few years there has been something of a revival
of interest in $Q$. The reasons for this include the following:
\begin{itemize}
\item Some understanding has been obtained into how $Q$ fits into the
QISM/quantum-groups picture of solvable lattice models \cite{PaGa92,BLZb,AF97,De99,Pr99,Pr00,BOU01,FR02}.
\item The discovery of the mysterious ODE/IM models correspondence - relating
  functional relations obeyed by the solutions and spectral determinants of certain
  ODEs to Bethe ansatz functional relations \cite{DT99a,BLZ99,DT99b}. 
\item The role of $Q$ in classical integrable systems as a generator of 
Backl\"und transformation has been understood in certain cases
(see \cite{Skl00} and references therein). 
\end{itemize}

In this paper, we are concerned with the first point. The key to the QISM
approach to solvable lattice models is to understand them in
terms of an underlying algebra $\cA$. The generators of $\cA$ are 
matrix elements $\cL^{ij}(z)$, where $i,j\in\{0,1\}$ (in the simplest
case) and $z$ is a spectral
parameter. The set of relations amongst the generators 
are given by the matrix relation
\be R(z/z') \cL_{1}(z) \cL_{2}(z') = \cL_{2}(z') \cL_{1}(z) R(z/z'),\lb{RLL}\ee
where $\cL_{1}(z)=\cL(z)\ot 1$, $\cL_{2}(z)=1\ot \cL(z)$, and $R(z)$ is a
$4\times 4$ matrix.

This QISM description was later refined
in terms of quantum groups. In this picture
$\cA$ is recognised as a quasi-triangular Hopf 
algebra (aka a quantum group). For the vertex models of the title,
the algebra $\cA$
is $\uq$. Families of $R$-matrices and $\cL$-operators are then all
given
in terms of representations of a universal $R$-matrix 
$\cR\in \uqbp\ot \uqbm$, where $U_q(b_\pm)$ are two Borel subalgebras
of $\uq$. The relevant $\uq$ representations are the 
spin-n/2 evaluation representations $(\pi^{(n)}_z,V^{(n)}_z)$ defined
in Section 3 (in this paper, a representation of an algebra $\cA$ is 
specified by a pair $(\pi,V)$, consisting of an $\cA$ module $V$ and the
associated map $\pi:\cA \to \End(V)$).
Then we have
\ben R(z/z')\equiv (\piz\ot \pi^{(1)}_{z'}) \cR, \quad \cL(z)\equiv(\piz \ot \id)
\cR,\een
and \mref{RLL} follows as a simple consequence of the Yang-Baxter
relation for $\cR$. 
More generally, we can define 
\be
\cT^{(n)}(z)\equiv\Tr_{\vzn} (\cL^{(n)}(z)), \quad \hb{where}
\ws \cL^{(n)}(z)\equiv (\pizn \ot \id) \cR\quad \hb{and}\ws n \in \Z
\, .\lb{tdefn}\ee
The $\cT^{(n)}(z)$ form a family of $\uqbm$ valued transfer matrices. 
The transfer matrix for a particular lattice model 
is given by choosing
a representation of one of the  $\cT^{(n)}(z)$'s over a particular `quantum space'.
For the homogeneous $N$ site 6-vertex model, the quantum space is the N-fold
tensor product $\vzo \ot \vzo \ot \cdots \vzo$, and the transfer 
matrix of the lattice model is 
\ben T^{(1)}(z)=(\pi^{(1)}_{1} \ot \pi^{(1)}_1 \ot \cdots \ot \pi^{(1)}_1) \cT^{(1)}(z).\een

Let us now consider how functional relations among the $\cT^{(n)}(z)$ 
arise.
The starting point is to note that tensor products of the $V_z^{(n)}$ have the following structure
\begin{prop}\lb{cp}
\ben 
 (a) \quad &&V_{zq^{n+1}}^{(n)} \ot V_z^{(1)} \quad \hb{has a unique
  proper} \ws \uq \ws \hb {submodule   }
V_{zq^{n+2}}^{(n-1)}, \hb{  and furthermore  }\\&&
V_{zq^{n+1}}^{(n)} \ot
V_z^{(1)}/\,V_{zq^{n+2}}^{(n-1)} \simeq V_{zq^n}^{(n+1)},\\[2mm]
 (b) \quad &&V_z^{(1)} \ot V_{zq^{n+1}}^{(n)} \quad \hb{has a unique
  proper} \ws \uq \ws \hb {submodule   }
V_{zq^{n}}^{(n+1)}, \hb{  and furthermore  }\\&&
V_z^{(1)}\ot V_{zq^{n+1}}^{(n)}/\,V_{zq^{n}}^{(n+1)} \simeq V_{zq^{n+2}}^{(n-1)}.\een
\end{prop}
\nin This proposition is a specialization of the more general tensor
product theorem due to Chari and Pressley \cite{ChPr91}.
Now, if $\Delta$ denotes the coproduct of $\uq$, then the following is a 
consequence of the defining properties of $\cR$:
\ben (\pi_{zq^{n+2}}^{(n)} \ot \pi_z^{(1)} \ot \id) (\gD\ot 1) \cR= (\pi_{zq^{n+2}}^{(n)} \ot \pi_z^{(1)} \ot \id) \, 
\cR_{13}\cR_{23}
= \cL_1^{(n)}(zq^{n+2})\cL_2^{(1)}(z).\een
If we take the trace of both sides of this equation over
$V_{zq^{n+1}}^{(n)} \ot V_z^{(1)}$, and use part (a) of Proposition
\ref{cp} as well as the property of the trace
given by Proposition \ref{trace} in order to rewrite the lhs,
 we arrive at the functional relation
\be \cT^{(n-1)}(zq^{n+2}) + \cT^{(n+1)}(zq^{n})=
\cT^{(n)}(zq^{n+2}) \cT^{(1)}(z).\lb{fr1}\ee
Similarly, by using part (b) of Proposition \ref{cp} we obtain
\be \cT^{(n-1)}(zq^{n+2}) + \cT^{(n+1)}(zq^{n})=  \cT^{(1)}(z) \cT^{(n)}(zq^{n+2}).\lb{fr2}\ee
Such functional relations and this approach to deriving them are
discussed in many places -
see for example \cite{KNS94}.

Baxter's $Q$-operator also obeys functional relations of a rather
similar form to \mref{fr1} and \mref{fr2} (see \cite{Bax82}), and 
in \cite{BLZb,AF97} the authors showed that it was possible to obtain 
such relations by constructing $Q$ in a manner similar to the above.
In analogy with \mref{tdefn}, they proposed constructing $Q$ operators
$Q_{\pm}(\gl)$ as
\be Q_{\pm}(\gl)=\Tr_{V_{\pm}(\gl)}\left(\,(\pi_{\pm}\ot \id)
  \cR\,\right) \,,\lb{af}\ee
where $(\pi_{\pm},V_{\pm}(\gl))$ were infinite-dimensional `q-oscillator' representations
of the Borel sub-algebra $\uqbp$ of $\uq$ \cite{AF97}.

In this paper, we consider more general infinite-dimensional
representations $(\pi_{(z,\us)},M(z,\us))$ of $\uqbp$, parameterized in
terms of a spectral parameter
$z$ and a vector $\us=(s_0,s_1,s_2)\in \C^3$.
Following \cite{BLZb,AF97}, we use them to define a $Q$-operator 
\ben \cQ(z,\us) = \Tr_{M(z,\us)}\big(\, (\pis\ot 1)\cR\, \big).\een 
We then consider tensor products of $M(z,\us)$ with $\vz$ and using the
result expressed in Proposition \ref{th1} go 
on to derive
generalized $T$-$Q$ relations 
\be \cT^{(1)}(z)\, \cQ(z,\us) =  \cQ(z,\us) \,\cT^{(1)}(z) = \cQ(zq^2,\us^+) + \cQ(zq^{-2},\us^-)\lb{tq0}.\ee
These relations 
involve the shifted vectors $\us^{\pm}=(q^{\pm 1} s_0, s_1, q^{\pm 2} s_2)$.
After a particular
specialization of the vector $\us$, we use the the appropriate
representation 
of $\cQ(z,\us)$ to construct an explicit form of the $Q$-operator for
the 6-vertex model (see \mref{explicit}).  This
construction works for all diagonal blocks of $Q(z,\us)$. 
Furthermore, the `spin zero' central block
of this operator coincides, up to an overall divergent factor, with 
Baxter's explicit expression for this block given by equation (101) of
\cite{Bax73aI} (Baxter's construction yields an explicit expression for this block only).

The layout of the paper is as follows: In Section 2, we define 
the infinite-dimensional representation $M(z,\us)$ of $\uqbp$ and 
give Proposition \ref{th1} concerning its tensor products with $\vz$.
In Section 3, we define a generalized $Q$-operator
$\cQ(z,\us)$ and derive the $T$-$Q$ relations \mref{tq0}. We show that
$\cT^{(1)}(z')$ and $\cQ(z,\us)$ commute. In Section 4, we give the
explicit form, $\bQ(z,\us)$, of our $Q$-operator for the 6-vertex model on 
a lattice with $N$ sites, and show 
how the coefficients arise in the $T$-$Q$ relations (i.e. the 
coefficients $\prod \limits _{i=1}^N\, \phi_1(z,\us,w_i)$ and $\prod \limits _{i=1}^N\, \phi_2(z,\us,w_i)$ in
\mref{tqu}). We also discuss the commutation relations of $\bQ(z,\us)$ and $\bQ(z',\us')$.
In Section 5, we give the explicit form \mref{explicit} of $\bQ(z,\us)$ for the 6-vertex
for a particular specialization of the parameter $\us$.
We give the connection with Baxter's explicit expression for the central block.
Finally, in Section 6, we make some observations about our construction
and discuss some possible avenues of work for the future.

\setcounter{equation}{0}
\section{Infinite-dimensional representations of $U_q(b_+)$}\lb{s2}

In this section, we define a level-zero representation $M(z,\us)$ of the
Borel subalgebra $U_q(b_+)$ of $\uqp$. 
We then consider the tensor products of $M(z,\us)$
with the spin-$1/2$ $\uqp$ evaluation module $V^{(1)}_z$. We also
give the restrictions of $\us$ for which $M(z,\us)$ reduces to a
q-oscillator representation.

\ssect{Definition of $M(z,\us)$}
First, let us recall the definition of $\uqp$ and $V^{(n)}_z$
(see, for example, \cite{chpr94} for an introduction to quantum affine algebras).
$\uqp$ is the associative algebra over $\C$ generated by the letters
$e_i,f_i,t_i,t^{-1}_i$, with $i\in \{0,1\}$, and with relations
\be
&&[e_i,f_j]=\delta_{i,j} \frac{t_i-\ti_i}{q-q^{-1}},\\
&&t_i e_i \ti_i = q^2 e_i, \quad t_i e_j \ti_i = q^{-2} e_j \quad (i\neq
j),\\
&&t_i f_i \ti_i = q^{-2} f_i, \quad t_i f_j \ti_i = q^{2} f_j \quad (i\neq
j),\\
&&e_i e_j^3 -[3] e_j e_i e_j^2 + [3] e_j^2 e_i e_j - e_j^3 e_i=0 \quad (i\neq j),\lb{Serre}\\
&&f_i f_j^3 -[3] f_j f_i f_j^2 + [3] f_j^2 f_i f_j - f_j^3 f_i=0 \quad
(i\neq j).\ee
We use the coproduct $\gD: \uqp \ra \uqp \ot \uqp$ given by
\ben
\gD(e_i) = e_i \ot 1 + t_i \ot e_i,\quad \gD(f_i) = f_i \ot t_i^{-1} + 1
\ot f_i,\quad \gD(t_i)=t_i\ot t_i.\een
Note, that the prime on $\uqp$ indicates that we are not including a
derivation in the definition. In this paper, we consider $\uqp$ and
its
representations at generic values of $q$ (i.e., $q$ is not a root of
unity).

The $\uqp$ evaluation module $V^{(n)}_z$, $n\in\Z_{>0}$, is defined in terms of basis
vectors $\vbn_j \ot z^m$ with $j\in\{0,1,\cdots,n\}$ and $m\in \Z$.
The $\uqp$ action is given by
\begin{equation}\begin{array}{lll}
&& e_1 (\vbn_j\ot z^m) = [j]\,(\vbn_{j-1} \ot z^m), 
 \quad f_1 (\vbn_j\ot z^m) = [n-j]\,(\vbn_{j+1} \ot z^m),\\[2mm]&&
t_1 (\vbn_j\ot z^m) = q^{n-2j}(\vbn_j\ot z^m),\\[1mm]
&& e_0 \sim (1\ot z)f_1,\quad f_0 \sim (1\ot z^{-1})\, e_1,
\quad  t_0\sim t_1^{-1}.\lb{act2}\end{array}\end{equation}
We use $(\pizn,\vzn)$ to denote the spin-$n/2$
evaluation representation consisting of the module $V^{(n)}_z$
and the associated map $\pizn:\uqp\ra \End(\vzn)$.

The algebra $\uqbp$ is defined as the Borel subalgebra of $\uqp$
generated by $e_1,e_0,t_1,t_0$, and $\uqbm$ is defined to be
the Borel subalgebra
generated by $f_1,f_0,t_1,t_0$.
Suppose we set out to 
define a $\uqbp$ module in terms of basis vectors $\ket{j}$, $j\in \Z$,
in the following way:
\be e_1 \ket{j} &=& \ket{j-1}, \quad e_0 \ket{j} = \gamma_j \ket{j+1},
\quad t_1 \ket{j}=s_0 q^{-2j}
\ket{j}, \lb{act}\ee
where $\gamma_j$ and $s_0$ are,  as yet, unknown coefficients (note that we can
always absorb an additional coefficient in the $e_1 \ket{j} = \ket{j-1}$
relation into a normalization of the basis vectors). Then for consistency
with the Serre relations \mref{Serre}, $\gamma_j$ must satisfy
\be \gamma_{j-3} - [3] \gamma_{j-2} + [3] \gamma_{j-1} - \gamma_j =0.\lb{djcond}\ee
The general solution of this recursion relation is 
\be \gamma_j= r+s_1 q^{2j} + s_2 q^{-2j} \lb{dj}\ee
where $r,s_1,s_2$ are arbitrary constants. Thus we can specify such a 
$\uqbp$ module by giving the four parameters $r,s_0,s_1,s_2$.
In fact, we choose to write $r$ in terms of $s_1,s_2$ and a new 
parameter $z$. We make the following definition:

\begin{dfn}\lb{mdef}
$M(z,\us)$ is a $\uqbp$ module specified in terms of basis vectors
$\ket{j}$, $j\in \Z$, $z\in \C \backslash \{0\}$ and a vector $\us=(s_0,s_1,s_2)\in \C^3$.
The $\uqbp$ action is given by
\ben e_1 \ket{j} &=& \ket{j-1}, \quad e_0 \ket{j} =
d_j(z,s_1,s_2) \ket{j+1},\quad 
t_1  \ket{j} = s_0 q^{-2j} \ket{j}, \quad t_0 \sim t_1^{-1}, \quad \hb{where}\\[3mm]
d_j(z,s_1,s_2) &\equiv &s_1s_2
\frac{(q-q^{-1})^2}{z}+\frac{z}{(q-q^{-1})^2}+s_1q^{2j}+s_2q^{-2j}. \een
\end{dfn}
\nin We use the notation $(\pis,M(z,\us))$ to indicate the representation
consisting of the $\uqbp$ module $M(z,\us)$
and the associated map $\pis:\uqbp\ra \End(M(z,\us))$.

\ssect{The tensor product structure}
Let us consider tensor products of $M(z,\us)$
and $V^{(1)}_z$ as $\uqbp$ modules. We have the following proposition:
\begin{prop}\lb{th1}

If $ \us^\pm\equiv (q^{\pm 1}s_0,s_1,q^{\pm 2}s_2)$ then
\ben
&(a) \quad &M(z,\us)\ot V^{(1)}_z \hb{  has a } \uqbp \hb{  submodule  } M(zq^2,\us^+),  \hb{
  and    }\\&&
\big( M(z,\us)\ot V^{(1)}_z\big)\, /M(zq^2,\us^+) \simeq M(zq^{-2},\us^-),\\[3mm]
&(b) \quad &V^{(1)}_z \ot M(z,\us) \hb{ has a } \uqbp \hb{ submodule } M(zq^{-2},\us^-), \hb{ 
and   } \\&&
 \big( V^{(1)}_z \ot M(z,\us)\big) /M(zq^{-2},\us^-) \simeq M(zq^2,\us^+).\een
\end{prop}
\begin{proof}
Let us first prove (a).
Define $A_j\equiv \big(  a_j \ket{j}\ot \vb_0 + \ket{j-1}\ot \vb_1\big)\in M(z,\us)\ot V^{(1)}_z$, where
the coefficient $a_j$ is as yet undetermined.
Clearly we have $t_1 A_j = s_0^+ q^{-2j} A_j$ where $s_0^+ = q s_0$.
The condition
that $e_1 A_j = A_{j-1}$ for all $j\in \Z$ is equivalent to
\be a_j+s_0 q^{2(1-j)} =a_{j-1}\, . \lb{se1}\ee
The condition that $e_0 A_j = \kappa_j A_{j+1}$ for some coefficient $\kappa_j$ and for all $j\in \Z$
is equivalent to
\be
&& a_j \, d_j(z,s_1,s_2) = a_{j+1}\, \kappa_j,\quad \hb{and}, \lb{se2}\\
&& a_j \, q^{2j} s_0^{-1} z + d_{j-1}(z,s_1,s_2)= \kappa_j, \lb{se3}
\ee
where the function $d_j(z,s_1,s_2)$ is specified in Definition \mref{mdef}.
Solving equations \mref{se1}-\mref{se3} gives
\be 
a_j &=& -\frac{ s_0 s_1 (1-q^2)}{zq^{2}}-\frac{s_0 q^{2(1-j)}}{(1-q^2)},\lb{c2}\\
\kappa_j &=&  d_j(zq^2,s_1^+,s_2^+),\ee
where $s_0^+, s_1^+$ and $s_2^+$ are the components of $\us^+$.
In this way, we have shown
that when \mref{c2} holds, $A_j$ are basis vectors
 of a submodule isomorphic to $M(zq^2,\us^+)$.

Now consider $B_j\equiv\ket{j+1}\ot \vb_0 \in  M(z,\us)\ot V^{(1)}_z$.
We immediately have that
\ben t_1 B_j = s_0 q^{-1} q^{-2j}B_j=s^-_0 q^{-2j}B_j\, ,\quad \hb{and}
\quad e_1 B_j = B_{j-1}.\een
It is also simple to establish that 
\ben e_0 B_j &=& z q^{2(j+1)} s_0^{-1} A_{j+2}+ d_j(zq^{-2},s_1^-,s_2^-) B_{j+1}.\een
Hence we have that
$\big( M(z,\us)\ot V^{(1)}_z\big)\, /M(zq^2,\us^+) \simeq M(zq^{-2},\us^-)$. 

The proof of (b) is very similar, the only significant differences are that
the analogue of the vector $A_j$ (which is now in the submodule $M(zq^{-2},\us^-)$) is
of the form 
\ben A_j= a_j \vb_0 \ot \ket{j+1}+ q^{j} \vb_1\ot \ket{j} \een for
some $a_j$,
and the analogue of $B_j$ (now in the quotient module $M(zq^2,\us^+)$) 
takes the form $q^{-j} \vb_0 \ot \ket{j}$.
\end{proof}

\ssect{The q-oscillator case}

Those $\uqbp$ representations on which either $(e_0e_1-q^2e_1e_0)$ or
$(e_1e_0-q^2e_0e_1)$ acts as a constant are referred to as q-oscillator 
representations \cite{Ku91}.
In terms of the action \mref{act}, these requirements
become either 
\be 
&&\gamma_{j-1}-q^2 \gamma_j=\hb{constant},\quad \hb{or}\lb{cd1}\\
&&\gamma_{j}-q^2 \gamma_{j-1}=\hb{constant}\lb{cd2}\ee
respectively.
Either of these conditions separately implies \mref{djcond}. The general
solutions of \mref{cd1} and \mref{cd2} are 
\ben
 \gamma_j&=&r + s_2 q^{-2j},\quad \hb{and} \\
 \gamma_j&=&r + s_1 q^{2j}\een
respectively, where $r,s_1,s_2$ are arbitrary constants.

Thus the specializations $M(z,s_0,0,s_2)$ and $M(z,s_0,s_1,0)$ are
both q-oscillator representations. Connecting with the notation
$V_{\pm}(\gl)$
notation of 
\cite{AF97}: $M(\gl,s_0,0,s_2)$ is a representation of
the type $V_+(\gl)$ and $M(\gl,s_0,s_1,0)$ is a 
representation of the type $V_-(\gl)$.\footnote{In \cite{AF97}, the 
notation $V_{\pm}(\gl)$ seems to refer originally to a class of 
representations; but in Appendix B, $V_+(\gl)$ refers to a 
specific representation, and in this case we have
$V_+(\gl)\simeq M(\gl,1,0,0)$.} The two operators $Q_{\pm}(\gl)$ of
\cite{AF97} (see our \mref{af}) are obtained by specializing our construction \mref{def1}
accordingly.

\setcounter{equation}{0}
\section{The Generalized $Q$-Operator}
In this section, we will construct a $\uqbm$
valued $Q$-operator $\cQ(z,\us)$ in terms of the universal $R$-matrix and
the $\uqbp$ module $M(z,\us)$ defined in the last
section. We will go on to show how generalized $T$-$Q$ relations arise
as a consequence of Proposition \ref{th1}. We will then consider
representations of $\cQ(z,\us)$.

\ssect{The operator $\cQ(z,\us)$}
We make use of the universal $R$-matrix of $\uq$, which we
denote by $\cR\in \uqbp\ot \uqbm$. The definition of $\cR$ can be
found in \cite{Dri86}; here we need only the properties 
\be &&(\gD \ot \id) \cR = \cR_{13}\cR_{23}, \quad (\id\ot \gD)
\cR = \cR_{13}\cR_{12},\lb{rprop1}\\
&&\cR_{12}\cR_{13} \cR_{23}=\cR_{23}\cR_{13}\cR_{12},\lb{yb} 
\ee
where as usual $\cR_{12} = \cR \ot \id$ etc.

Then using the representations defined in Section \ref{s2},
we make the following definitions:
\ben \cL(z)&=&(\piz \ot \id) \cR \in \End(V^{(1)}_z)\ot \uqbm,\\[2mm]
\cW(z,\us)&=&(\pis \ot \id) \cR \in \End(M(z,\us))\ot \uqbm.
\een
By taking the trace we go on to define 
\be 
\cT(z)&=& \Tr_{V_z^{(1)}}(\cL(z)) \in  \uqbm,\lb{def0}\\
\cQ(z,\us)&=&  \Tr_{M(z,\us)}(\cW(z,\us))\in \uqbm. \lb{def1}
\ee
The operators $\cL(z)$ and $\cT(z)$ are the familiar monodromy
matrix and transfer matrix of the QISM
(although these terms are perhaps more commonly reserved for
representations of these algebraic objects on particular quantum
spaces). The operator $\cQ(z,\us)$ is our generalized $Q$-operator.

\ssect{$T$-$Q$ relations}
Our starting point in the derivation of $T$-$Q$ relations is the
following simple proposition:
\begin{prop}{\lb{trace}}
If $\cA$ is an associative algebra, $X\in \cA$, and A,B,C are finite-dimensional $\cA$ 
modules which form an exact sequence
$0\ra B\ra A\ra C \ra 0$, then
$\Tr_A(X)= \Tr_B(X) + \Tr_C(X)$.
\end{prop}

\nin A proof is given in Appendix A.
To proceed, let us consider the expression
\be
(\pis \ot \pi_z^{(1)} \ot \id) (\gD\ot 1) \cR \in \End(M(z,\us)) \ot
\End(V^{(1)}_z) \ot \uqbm.\lb{ex1}\ee
Using the first property in \mref{rprop1}, we arrive at
\be (\pis \ot \pi_z ^{(1)}\ot \id) (\gD\ot 1) \cR= (\pis \ot \pi_z^{(1)} \ot \id) \, 
\cR_{13} \cR_{23}
= \cW_1(z,\us) \, \cL_2(z).\lb{eq1}\ee

Suppose we assume that Proposition \ref{trace} also holds for the
{\it infinite-dimensional} $\uqbp$ modules involved in the exact sequence 
\ben 0 \ra M(zq^2,\us^+) \ra M(z,\us) \ot V^{(1)}_z \ra M(zq^{-2},\us^-) \ra 0,\een
whose existence is equivalent to part (a) of
Proposition \ref{th1}. In this case, we will have 
\be \Tr_{M(z,\us)\ot V_z^{(1)}} (\gD(X))=\Tr_{M(zq^2,\us^+)}(X) +
 \Tr_{M(zq^{-2},\us^-)}(X),\lb{eq2}\ee
for $X\in \uqbp$.
Taking the trace over $M(z,\us)\ot V_z^{(1)}$ of both sides of \mref{eq1}
and using \mref{eq2} to rewrite the lhs (as well as the definitions
\mref{def0}
and \mref{def1})  yields
\ben \cQ(zq^2,\us^+) + \cQ(zq^{-2},\us^-) = \cQ(z,\us) \, \cT(z) \in
\uqbm.\een
A similar argument, which now relies on part (b) of Proposition \ref{th1}
gives
\ben \cQ(zq^{-2},\us^-) + \cQ(zq^2,\us^+) =  \cT(z) \,\cQ(z,\us)\in
\uqbm .\een
Thus we arrive at the $T$-$Q$ relations
\be \cT(z)\, \cQ(z,\us) =  \cQ(z,\us) \, \cT(z) = \cQ(zq^2,\us^+) +
\cQ(zq^{-2},\us^-)\, .\lb{tq1}\ee

In Sections 4.3 and 5, we discuss the
meaning of the infinite-dimensional trace involved in the definition
of $\cQ(z)$, and describe various checks of the $T$-$Q$ relations \mref{tq1} that arise as
a consequence of the assumption about the validity of the extension 
of Proposition \ref{trace}.

\ssect{Commutation relations}
The Yang-Baxter relation \mref{yb} is an equality between elements of $\uqbp\ot \uqp \ot \uqbm$. Acting on both sides with $\pis \ot \pi^{(1)}_{z'} \ot \id$ (where ${z'}\in \C$ is
arbitrary)
gives
\ben
&&W(z,\us;{z'})\,  \cW(z,\us ) \, \cL({z'}) =\cL({z'}) \,  \cW(z,\us )\, W(z,\us;{z'})
\quad \in \End_{M(z,\us)} \ot \End_{V_{z'}^{(1)}} \ot \,\uqbm,\een
where $W(z,\us;{z'}) \equiv (\pis \ot \pi^{(1)}_{z'} )\cR$.
Multiplying on the right by $W(z,\us;{z'})^{-1}$, taking the trace over $M(z,\us)\ot V_{z'}^{(1)}$, and using the definitions
\mref{def0} and \mref{def1} we obtain the commutation relations 
\be [\cQ(z,\us), \cT({z'})]=0.\lb{com1}\ee

It is an obvious next step to attempt to repeat this argument and 
act with $\pis \ot \pi_{({z'},\us')} \ot \id$ on \mref{yb} in order
to derive commutation relations involving $\cQ(z,\us)$ and $\cQ(z',\us')$.
However such an argument {\it fails} for the simple reason that 
$M(z',\us')$ is only a $\uqbp$ module and not a $\uqp$ module.
In fact, as discussed in detail in Section 4.4,
$\cQ(z,\us)$ and $\cQ(z',\us')$ 
do {\it not} commute for general $\us$ and $\us'$.

\ssect{Representations of $\cQ(z,\us)$}
We have constructed both $\cT(z)$ and $\cQ(z,\us)$ as $\uqbm$ valued
objects. Constructing the operators corresponding to a particular
lattice model simply involves choosing a representation of $\uqbm$
(i.e., choosing a $\uqbm$ module $V_{qu}$ and the associated map $\pi_{qu}:\uqbm \ra
\End(V_{qu})$\,).
Such a representation is referred to as the quantum space in the 
language of the QISM. 
If we define
\ben 
T(z) \equiv \pi_{qu}(\cT(z))\in \End(V_{qu})\quad \hb{and}\quad
Q(z,\us) \equiv \pi_{qu}(\cQ(z,\us))\in \End(V_{qu}),\een
it then follows from \mref{tq1} and \mref{com1} that we have the $T$-$Q$ relations
\be T(z)\, Q(z,\us) =  Q(z,\us) \, T(z) = Q(zq^2,\us^+) +
Q(zq^{-2},\us^-)\lb{tq2},\ee 
and the commutation relations 
\ben [Q(z,\us), T(z')]=0. \lb {com2}\een

If we choose our quantum space to be equal to
$V_1\ot V_2 \ot \cdots \ot V_N$, as will be the case for lattice models, then it follows 
from \mref{rprop1} that we
have 
\ben 
T(z) &=& \Tr_{V_z^{(1)}}\big(L_N(z)  L_{N-1}(z)   \cdots
L_1(z)\big)\quad \hb{and}\\
Q(z,\us) &=& \Tr_{M(z,\us)}\big(W_N(z,\us)    W_{N-1}(z,\us)   \cdots
  W_1(z,\us)\big), \quad \hb{where},\\[3mm]
 L_i(z) &\equiv & (\id \ot \pi_{V_i})\, \cL(z) \in \End(V_z^{(1)}) \ot \End(V_i),
 \quad \hb{and}\\
 W_i(z,\us) &\equiv &(\id \ot  \pi_{V_i})\, \cW(z,\us) \in \End(M(z,\us)) \ot \End(V_i).\een

Let us comment on the connection between our $T$-$Q$ relations \mref{tq2}
and
Baxter's $T$-$Q$ relations \cite{Bax82}. Two differences are immediately apparent. Firstly, 
there are more parameter in our relation; $Q(z,\us)$ depends upon
the spectral parameter $z$ and $\us=(s_0,s_1,s_2)\in\C^3$ as well 
as the parameter $q$ in the universal
$R$-matrix and all parameters associated with the quantum
space $V_{qu}$.
In the Section 5, we compute our $Q$-matrix explicitly for the
6-vertex model, and show how with a particular specialization it is related to 
Baxter's $Q$-matrix. Secondly, the coefficients multiplying the
$Q(zq^2,\us^+)$ and $Q(zq^{-2},\us^-)$ on
the rhs of \mref{tq2} are $1$, unlike in Baxter's case.
This is a trivial point associated with the 
the normalization  of the $Q$-matrix that we have used in this section.
The coefficients will reappear in Section  
4.2, when we choose to normalize our $Q(z,\us)$ in a more practical way.

\setcounter{equation}{0}
\section {Properties of $Q$ for the 6-Vertex Model}

In this section, we will consider the operator $W_i(z,\us)$ 
when the
$V_i$ appearing in the quantum space \,$V_1\ot V_2 \ot \cdots
V_N$ \,
is equal to $V_{w _i}^{(1)}$, i.e.,  in the 6-vertex
model case. We will determine the action of $W_i(z,\us)$
on the space $M(z,\us)\otimes
V_{w _i}^{(1)}$ up to a multiplicative constant, and show how the 
coefficients appearing on the rhs of the $T$-$Q$ relation depend on this
normalization factor.
Finally, we will address the problem of the
commutativity of $Q$-operators.

\subsection {Definition of $W_i$ and $Q$ for the 6-vertex model}

Let us consider the operator
\begin{equation}
 W(z,\us; w)\equiv (\pi
_{(z,\us)}\otimes \pi _{w}^{(1)}){\cal R} \,\in {\mbox {End}} (M(z,\us))\otimes {\mbox {End}}(
V_{w}^{(1)}) .\label {S}
\end{equation}
$W(z,\us; {w}_i)$ is the operator $W_i(z,\us)$ associated with the 6-vertex
model with local inhomogeneity parameter $w_i$.

In order to determine $W(z,\us; w)$ explicitly, definition (\ref
{S})  is rather inconvenient, because it involves the unwieldy universal $R$
matrix. It is easier to construct an operator  $\bW(z,\us;w)\in
{\mbox {End}} (M(z,\us))\otimes {\mbox {End}}(V_{w}^{(1)})$
which satisfies the properties
\begin{eqnarray}
\bW(z,\us;w)(\pi _{(z,\us)}\otimes \pi _{w}^{(1)}) \Delta
(t_i)&=&(\pi _{(z,\us)}\otimes \pi _{w}^{(1)}) \Delta ^\prime
(t_i)\bW(z,\us;w) \, , \lb{copr1}\\ \bW(z,\us;w)(\pi
_{(z,\us)}\otimes \pi _{w}^{(1)}) \Delta (e_i)&=&(\pi
_{(z,\us)}\otimes \pi _{w}^{(1)}) \Delta ^\prime
(e_i)\bW(z,\us;w) \, . \lb{copr2}
\end{eqnarray}
Since $M(z,\us)\otimes
V_{w}^{(1)}$ is an irreducible $\uqbp$ module for generic $z$, $\us$
and $w$, 
it follows that an operator satisfying
properties (\ref {copr1}, \ref {copr2}) is unique up to a
multiplicative constant. 
By definition, $W(z,\us;w)$ also satisfies (\ref {copr1}, \ref {copr2})
and so will be proportional to
$\bW(z,\us;w)$.

Let us solve (\ref {copr1}, \ref {copr2}). 
Firstly,   relation
\mref{copr1} requires that $\bW(z,\us; w)$ must be of the form
\begin{eqnarray}
\bW(z,\us;w)\, |j>\otimes \, v_0^{(1)} &=&\alpha _{j,0}\,
|j>\otimes \, v_0^{(1)}+\beta _{j,0}\, |j-1>\otimes \, v_1^{(1)} \, ,
\lb{Sv0}\\ \bW(z,\us;w)\, |j>\otimes \, v_1^{(1)} &=&\alpha
_{j,1}\, |j>\otimes \, v_1^{(1)}+\beta _{j,1}\, |j+1>\otimes \,
v_0^{(1)}\, , \lb{Sv1}
\end{eqnarray}
where $\alpha _{j,0}$, $\alpha _{j,1}$, $\beta _{j,0}$ and $\beta
_{j,1}$ are arbitrary coefficients. 
Then, 
\mref{copr2} is satisfied if and only if
\be \beta _{j-1,0}&=&q^{-1}\beta _{j,0} \nonumber \\
\alpha _{j-1,0}&=&q\alpha _{j,0}+\beta _{j,0} \nonumber \\ 
\alpha
_{j-1,1}+s_0q^{-2j}\beta _{j,0}&=&q^{-1}\alpha _{j,1} \nonumber \\
\beta _{j-1,1}+s_0q^{-2j}\alpha _{j,0}&=&\alpha _{j,1}+q\beta _{j,1}
\nonumber \\
 d_j(z,s_1,s_2)\alpha _{j+1,0}+s_0^{-1}q^{2j}w \beta
_{j,1}&=&q^{-1}d_j(z,s_1,s_2)\alpha _{j,0} \nonumber \\ 
d_j(z,s_1,s_2)\beta
_{j+1,0}+s_0^{-1}q^{2j}w \alpha _{j,1}&=&w \alpha
_{j,0}+qd_{j-1}(z,s_1,s_2)\beta _{j,0} \nonumber \\ 
d_j(z,s_1,s_2) \alpha _{j+1,1}&=&q \,d_j(z,s_1,s_2)
\alpha _{j,1}+w \beta _{j,1} \nonumber \\
 d_j(z,s_1,s_2) \beta_{j+1,1}&=&q^{-1}d_{j+1}(z,s_1,s_2)\beta
 _{j,1},\nn
\ee
where the function $d_j(z,s_1,s_2)$ is specified in Definition \ref{mdef}.
These equations have the general solution  
\begin{equation}\begin{array}{lll} 
\alpha _{j,0}&=&
 \left(
\frac{s_2}{w}q(1-q^2)q^{-j}-\frac{q^j}{q-q^{-1}}\right)\rho, \quad 
\alpha _{j,1}= \left( \frac{s_0s_1}{w} q(q^{-2}-1)q^{j}- 
s_0\frac{q^{-j}}{q-q^{-1}})\right)\rho \\ 
 \beta _{j,0}&=& q^j \rho, \quad
 \beta _{j,1}= s_0\frac {q^{-j}}{w}\,d_j(z,s_1,s_2) \rho,  \lb{a0}\end{array}\end{equation}
where $\rho$ is an arbitrary constant. 

Let us choose the normalization constant $\rho$ to be an as yet 
unspecified function $\rho(z,\us,w)$, and define the $Q$-operator
in terms of $\bW(z,\us,w)$ (given by \mref{Sv0}, \mref{Sv1} and
\mref{a0}\,) by
\begin{equation}
\bQ(z,\us)={\mbox {Tr}}_{M(z,\us)}\left(\bW(z,\us;{w}_N) \bW(z,\us;{w}_{N-1}) \ldots  \bW(z,\us;{w}_1)\right) \, . \lb{Qdef}
\end{equation}

\subsection{Normalized $T$-$Q$ relations for the 6-vertex model}
Since we are dealing with the 6-vertex model, we have to consider the
Lax operator $L_i(z)$ of Section 3 in the case when the quantum space
is 
$V_i=V_{w _i}^{(1)}$.  
This is given by the matrix
\begin{equation}
R(z/w_i)=(\pi _{z}^{(1)} \otimes \pi _{w_i}^{(1)}){\cal R} \, , \lb{Rmat2}
\end{equation}
acting on $V_{z}^{(1)} \otimes V_{w_i}^{(1)}$. 
The matrix $R(z/w)$
is proportional to the normalised matrix $\bR(z/w)$
given by
\begin{equation}\begin{array}{lll}
\bR(z/w)\, v_0^{(1)} \otimes v_0^{(1)} &=& v_0^{(1)} \otimes v_0^{(1)} \\[2mm]
\bR(z/w)\, v_0^{(1)} \otimes v_1^{(1)} &=& \frac {q\left (1-\frac {z}{w}\right )}{1-q^2 \frac {z}{w}}v_0^{(1)}\otimes v_1^{(1)} + \frac {\frac {z}{w}(1-q^2)}{1-q^2 \frac {z}{w}}v_1^{(1)}\otimes v_0^{(1)}\\[2mm]
\bR(z/w)\, v_1^{(1)} \otimes v_0^{(1)} &=& \frac {q\left (1-\frac {z}{w}\right )}{1-q^2 \frac {z}{w}}v_1^{(1)}\otimes v_0^{(1)} + \frac {1-q^2}{1-q^2 \frac {z}{w}}v_0^{(1)}\otimes v_1^{(1)} \\[2mm]
\bR(z/w)\, v_1^{(1)} \otimes v_1^{(1)} &=& v_1^{(1)} \otimes v_1^{(1)} \, . \label{Rrep}
\end{array}\end{equation}
We define the normalized transfer matrix of the 6-vertex model by
\begin{equation}
\bT(z)={\mbox {Tr}}_{V_{z}^{(1)}}\left(\bR(z/w _N) \bR(z/w _{N-1})
  \ldots  \bR(z/w _1)\right)\, . \lb{tdef}
\end{equation}

We are going to derive the  coefficients in the  $T$-$Q$ relation
associated with  the   above  normalization   of   $\bT(z)$  and   $\bQ(z,\us)$.
As a preliminary, let us introduce a little more notation: we use
$\ket{j}^\pm$ to denote the basis vectors in $M(zq^{\pm 2},\us^\pm)$;
$\iota$ to 
denote
the embedding 
\ben
\iota:M(zq^2,\us^+) &\to&M(z,\us)\ot \vz\\
\ket{j}^+ &\goto&A_j;
\een
and $\pi$ to denote the projection
\ben 
\pi:M(z,\us)\ot \vz &\to&M(zq^{-2},\us^-)\\
     B_j &\goto&\ket{j}^-.\een
$A_j$ and $B_j$ are as defined in the proof of Proposition \ref{th1}.

The  coefficients of
the $T$-$Q$ relation  are then obtained from the action of 
$\bW(z,\us;   w)    \bR(z/w)$    on   the
space $M(z,\us) \otimes  V_{z}^{(1)} \otimes  V_{w}^{(1)}$.
By direct calculation we find
\be   \bW_{13}(z,\us;   w)
\bR_{23}(z/w)  A_j \otimes \,  v_0^{(1)}&=&\phi _1(z,\us,w)
[\alpha^{+}_{j,0}\,  A_j\otimes   \,   v_0^{(1)}  +   \beta  ^+   _{j,0}\,
A_{j-1}\otimes \, v_1^{(1)} ] ,\quad\quad \lb{sr1} \\[2mm]
\bW_{13}(z,\us;
w)\bR_{23}(z/w) A_j  \otimes  \, v_1^{(1)}&=&\phi  _1(z,\us,
w)[\alpha ^+ _{j,1}\, A_j\otimes \, v_1^{(1)} + \beta ^+ _{j,1}\,
A_{j+1}\otimes \, v_0^{(1)} ] \lb{sr2} , \\[2mm]
\bW_{13}(z,\us;
w)\bR_{23}(z/w) B_j  \otimes  \, v_0^{(1)}&=&\phi  _2(z,\us,
w)[\alpha  ^{-}  _{j,0}\,  B_j\otimes  \, v_0^{(1)}+  \beta  ^{-}
_{j,0}\, B_{j-1}\otimes \, v_1^{(1)} ] , \lb{sr3} \\[2mm]
 \bW_{13}(z,\us;
w)\bR_{23}(z/w) B_j  \otimes  \, v_1^{(1)}&=&\phi  _2(z,\us,
w)  [\alpha ^{-} _{j,1}\,  B_j\otimes \,  v_1^{(1)} +  \beta ^{-}
_{j,1} \, B_{j+1}\otimes \, v_0^{(1)} ]  \\[2mm]
 &\quad\quad +&\frac {z(1-q^2)}{w-q^2z}    \beta   _{j+1,0}   \,
A_{j+1}\otimes        \,        v_1^{(1)}  \nn \\    &\quad\quad+&
\frac{z(1-q^2)}{w-q^2z}   \alpha   _{j+1,0}   \,
A_{j+2}\otimes \, v_0^{(1)} , \lb{sr4} \ee where 
\be\nn \\[-17mm]
\phi _1(z,\us,w)\equiv \frac {\rho(z,\us,w)}{\rho(zq^2,\us^+,w)}
\frac {w-z}{w-q^2z} \,  , \quad \phi _2(z,\us,w) 
\equiv \frac {\rho(z,\us,w)}{\rho(zq^{-2},\us^-,w)} \,q \, . \label{phi}
\ee
$\alpha _{j,0}^\pm$, $\alpha _{j,1}^\pm$, $\beta _{j,0}^{\pm}$, $\beta
 _{j,1}^{\pm}$ are the coefficients (see relations \mref{Sv0}-\mref {a0}\,)
 specifying the action of $\bW(z,\us^\pm; w)$ on $M(z,\us^\pm)\otimes V_{w}^{(1)}$.

An immediate consequence of \mref{sr1}-\mref{sr4} is that we have
\ben
\bW_{13}(z,\us;   w)\bR_{23}(z/w) \iota(\ket{j}^+)\ot v^{(1)}_\ep &=& 
\phi _1(z,\us,w) (\iota\ot 1) \bW(zq^2,\us^+; w)\, \ket{j}^+\ot v^{(1)}_\ep\\
(\pi\ot 1) \bW_{13}(z,\us;   w)\bR_{23}(z/w)  B_j  \otimes  \, v_\ep^{(1)}&=&
\phi _2(z,\us,w)  \bW(zq^{-2},\us^-; w)\, \ket{j}^-\ot v^{(1)}_\ep,
\een
for $\ep\in\{0,1\}$. It then follows (by the appropriate modification of the
argument in the Proof in Appendix A by the factors $\phi_1$ and $\phi_2$) that
\ben 
\Tr_{M(z,\us)\ot \vz}\left(\bW_{13}(z,\us;w)\bR_{23}(z/w)\right)
&= &\phi _1(z,\us,w)\, \Tr_{M(zq^2,\us^+)}\left(\bW(zq^2,\us^+;w) \right) \\
&+ &\phi _2(z,\us,w) \Tr_{M(zq^{-2},\us^-)}\left(\bW(zq^{-2},\us^-;w)
\right).\een
This is the normalized $T$-$Q$ relation associated with a quantum space
$V_w^{(1)}$. More generally, a modification of the above argument
to the case when the quantum space is $V_{w_1}^{(1)}\ot V_{w_2}^{(1)}
\cdots V_{w_N}^{(1)}$ gives the normalized $T$-$Q$ relations
\begin{equation}
\bQ(z,\us )\,\bT(z)=\bT(z)\,\bQ(z,\us)=\left ( \prod _{i=1}^N \phi
 _1(z,\us,w _i) \right )\,
 \bQ(zq^2,\us ^+)
+\left ( \prod _{i=1}^N \phi _2(z,\us,w _i)\right ) \, \bQ(zq^{-2},\us ^-) \, . \label{tqu}
\end{equation}

\subsection{Some comments on the definition of $Q$}
We have defined our generalised $Q$-operator for the 6-vertex model
by formula (\ref {Qdef}). Let us make some comments on this definition.
First of all, we notice that, due to the charge conservation
property (\ref {copr1}), the
 operator (\ref {Qdef}) is a block-diagonal matrix, each block connecting vectors of $V_{{w}_1}^{(1)}\otimes \ldots \otimes V_{{w}_N}^{(1)}$ containing the same number of $v_0^{(1)}$. 
Next, we remark that
it follows from the explicit expression for $\bW(z,\us;w)$ given by
\mref{Sv0}-\mref{a0},
 that matrix elements of the $Q$-operator (\ref {Qdef}) will have the following form:
\ben
\sli _{k=0}^N a_k(z,\us,  w_1,\cdots,w_N, q)\, \delta (q^k) \, , 
\een
where $a_k(z,\us,w_1,\cdots,w_N,q)$ is a product of the
$\rho(z,\us,w_i)$ normalization factors with a rational function
of $z,\us,w_1,\cdots,w_N,q$\,, and $\delta (q^k)\equiv\sum_{j\in \Z}
q^{k\,j}$. When $k\neq 0$, $\delta(q^k)$ is a formal series; but clearly
some care is needed in interpreting the meaning of $\delta(q^k)$ when $k=0$\,!
The situation becomes clearer if we restrict
the case $s_1=s_2=0$: with this restriction this $\delta(q^0)$ appears only as an overall multiplicative
factor in the 
central block of $Q$ connecting $N/2$ vectors $v_0^{(1)}$, and thus
cancels from both sides of the $T$-$Q$ relation. Moreover, Baxter has an
explicit expression for the central block of his $Q$-operator \cite{Bax73aI},
and in Section 5  we establish an equality between the central block
of our
$Q$ 
(with the $s_1=s_2=0$ restriction and $\delta(q^0)$ replaced by
1) and Baxter's operator. 

This result leads us to conjecture that the $T$-$Q$  relations  \mref{tqu}
hold if we simply replace
$\delta(q^0)$, wherever it appears in the matrix elements of
$\bQ(z,\us)$,
by a constant (1 say).
As further support to this conjecture, we have checked explicitly in the case when $N=2$ and 
 $\us$ is generic, that the
 $T$-$Q$ relation holds independently of the value assigned to 
$\delta(q^0)$. 
 
\subsection{Commutation relations of $Q$ for the 6-vertex model}
In Section 3 we have constructed a generalised $Q$-operator
 $\cQ(z,\us)$ satisfying the $T$-$Q$ relation \mref{tq1} and commuting with
${\cal T}(z')$. We  have already remarked in Section  3.3 that, unlike
 the  commutativity with  ${\cal T}(z')$,  the commutativity  of ${\cal
 Q}(z,\us)$  with   $\cQ(z',\us')$ cannot  be  shown   by  general  algebraic
 arguments. Therefore,  in order  to deal with  this fact, we  have to
 choose particular representations for the quantum space. We 
 will  focus on the 6-vertex  model, for which  the $Q$-operator is
 given by $\bQ(z,\us)$ (\ref {Qdef}).

Let us  start by  saying that explicit  calculations performed  in the
case $N=2$  show that $\bQ(z,\us)$ and  $\bQ(z^\prime,\us ^\prime)$ do
{\it not}  commute  for  general   values  of  the  parameters  $z,\us$  and
$z^\prime,\us ^\prime$.  On the  other hand, we will show in the next
section
that such  commutativity holds for general $N$ when we restrict to the case
\begin{equation}
s_1=s_2=s_1^\prime=s_2^\prime=0 \, . \lb{case}
\end{equation}


\setcounter{equation}{0}
\section {An Explicit Form of $Q$ for the 6-Vertex Model}
 
In this section, we give a simple explicit expression for our $Q$-operator
(\ref {Qdef}) for general $N$, when we make the specialization $s_1=s_2=0$.
We shall find that this expression is related to equation (101) of
\cite{Bax73aI} (which is an expression for the central block of the
Q-matrix when the number of lattice sites $N$ is even). 

First of all, let us define matrix elements of $\bW(z,\us,w)$
and $\bQ(z;s)$ of Section 4.1. In this section, we will use the notation
$v^{(1)}_{+ (-)}$ in place of $v^{(1)}_{0 (1)}$ in order to facilitate
comparison with \cite{Bax73aI}. We define matrix elements by
\ben \bW(z,\us,w)\ket{j}\ot v^{(1)}_{\gb} &=&
\sli_{\ga} \bW(z,\us,w)^{j\,\gb}_{j'\,\ga} \ket{j}\ot
v^{(1)}_{\ga}, \quad \hbox{where} \ws j'=j+(\ga-\gb)/2,\\
\bQ(z,\us)(v^{(1)}_{\beta_1}\ot v^{(1)}_{\beta_2}\ot \cdots \ot
v^{(1)}_{\beta_N})
&=&\sli_{\ga_1,\ga_2, \cdots, \ga_N}
\bQ(z,\us)^{\beta_1,\beta_2,\cdots,\beta_N}_{\ga_1,\ga_2,
\cdots,\ga_N} (v^{(1)}_{\ga_1}\ot v^{(1)}_{\ga_2}\ot \cdots \ot
v^{(1)}_{\ga_N}).
\een
Fixing $w_1=\cdots=w_N=w$, it then follows from \mref{Qdef} that we have
\begin{equation}
\bQ(z,\us)_{\alpha _1 \ldots \alpha _N}^{\beta _1 \ldots 
 \beta _N}=\sum _{j\in\Z}\bW(z,\us;w)_{\, \, \,
 j,\alpha _N}^{j_{N-1},\beta _N}  \bW(z,\us;w)_{j_{N-1},\alpha
 _{N-1}}^{j_{N-2},\beta _{N-1}} \ldots \bW(z,\us;w)_{j_1,\alpha _1}^{j,\beta _1}\, .\lb{qu1}
\end{equation}
where
$j_k=j+\half(\ga_1+\ga_2+\cdots+\ga_{k-1})-\half(\gb_1+\gb_2+\cdots+\gb_{k-1})$.

Now let us consider the matrix elements 
$\bW(z,\us,w)^{j\,\gb}_{j'\,\ga} $  given by \mref{a0},
in the special case when $s_1=s_2=0$. In this case, we have 
\be 
\bW(z,\us,w)^{j\,+}_{j\,+} &= -\frac{\rho\, q^j}{q-q^{-1}},\quad
\bW(z,\us,w)^{j\,-}_{j\,-} &= -\frac{s_0 \,\rho \,q^{-j}}{q-q^{-1}},\lb{ade1}\\[2mm]
\bW(z,\us,w)^{j\,+}_{j-1\,-} &= \rho \, q^j \quad
\bW(z,\us,w)^{j\,-}_{j+1\,+} &= \frac{z}{w}\, \frac{s_0 \, \rho \,q^{-j} }{(q-q^{-1})^2}.\lb{ade2}
\ee
It a simple consequence of charge conservation
(i.e. the property that $\bW(z,\us,w)^{j\,\gb}_{j'\,\ga}$ is
      only non-zero for $j'=j+(\ga-\gb)/2$) that whenever we have a
      contribution to \mref{qu1} of the form
      $\bW(z,\us,w)^{\cdot\,+}_{\cdot\,-}$
we must also have one of the form
$\bW(z,\us,w)^{\cdot\,-}_{\cdot\,+}$. So when computing \mref{qu1}, we 
can equally use the more symmetric expressions 
\ben
\bW(z,\us,w)^{j\,+}_{j-1\,-} &= 
-\left(\frac{z}{w}\right)^\half \frac{ \sqrt{s_0} \, \rho \, q^{j} }{ q-q^{-1}},\quad 
\bW(z,\us,w)^{j\,-}_{j+1\,+} &= 
-\left(\frac{z}{w}\right)^\half \frac{ \sqrt{s_0} \, \rho \, q^{-j} }{ q-q^{-1}}.\lb{ade3}
\een
Then, if we define $\tilde{\rho}=-\frac{\sqrt{s_0} \rho}{q-q^{-1}}$,
we can write both \mref{ade1} and \mref{ade2} as
\be
\bW(z,\us,w)^{j\,\gb}_{j'\,\ga}= 
\left(\frac{z}{w}\right)^{(1-\ga \gb)/4} \, s_0^{-(\ga+\gb)/4}
 \, \tilde{\rho} \, q^{j\gb}.\lb{nice}\ee

Using the form \mref{nice}, it is then straightforward to compute a
simple expression for \mref{qu1}. The matrix is block diagonal,
consisting of blocks for which 
$\ga_1+\ga_2+\cdots\ga_N=\gb_1+\gb_2+\cdots\gb_N=n$, where
$n\in\{-N,-N+2,\cdots,N\}$. The block labelled by $n$ in this way
has the form
\be 
\bQ(z,\us)_{\alpha _1 \ldots \alpha _N}^{\beta _1 \ldots 
 \beta _N}=
s_0^{-n/2} \delta(q^n) \left(\frac{z q}{w}\right)^{N/4} \tilde{\rho}^N
\, q^{\frac{1}{4}\sli_{i<j}(\beta_j \alpha_i - \alpha_j \beta_i)}
  \left(\frac{w}{zq}\right)^{\frac{1}{4}\sli_i\ga_i\gb_i},\lb{explicit}
\ee
where the function $\delta(q^n)=\sli_{j\in\Z} q^{n j}$ arises from
sum over $j$ in expression \mref{qu1}.

Now, let us compare \mref{explicit} to Baxter's Q-matrix given 
by equation (101) of \cite{Bax73aI} as:
\be
 [Q_{{Bax}}(v)]_{\alpha _1 \dots  \alpha _N}^{\beta _1 \ldots  \beta _N}=\exp \left [ \frac {1}{2}i\eta \sum _{k=1}^N\sum _{j=1}^{k-1}(\alpha _j \beta _k - \alpha _k \beta _j)+\frac {1}{2}iv \sum _{j=1}^N \alpha _j \beta _j \right ] \, . \lb{Qbax}
\ee Clearly, if we identify
\be q=\exp(2 i \eta ), \quad \frac{w}{zq}=\exp(2iv),\lb{identify1}\ee
and choose the arbitrary normalization function $\tilde \rho$ to be
\be \tilde{\rho}=\tilde{\rho}(z,\us,w)=(zq/w)^{-\frac{1}{4}},\lb{rhores}\ee
then we have 
\be 
\bQ(z,\us)_{\alpha _1 \ldots \alpha _N}^{\beta _1 \ldots 
 \beta _N}= s_0^{-n/2} \delta(q^n) [Q_{{Bax}}(v)]_{\alpha _1
 \dots  \alpha _N}^{\beta _1 \ldots  \beta _N}.\lb{baxform}\ee
Let us emphasize, that whereas \mref{Qbax} is derived in 
\cite{Bax73aI} as an expression for the $n=0$ block of the Q-matrix,
our expression \mref{explicit} or \mref{baxform} is valid for 
all blocks (the $n=0$ case is discussed below).
It is also valid for $N$ both even and odd.
Also note that the choice \mref{identify1} is the
one required in order to identify (up to a normalization and gauge
transformation) our 6-vertex model R-matrix
$\bR(z/w)$ with Baxter's R-matrix given as a function of $v$ and
$\eta$.\footnote{The paper \cite{Bax73aI} primarily concerns the 8-vertex model, but the
explicit formula (101) is obtained in the 6-vertex model 
limit.}

Let us now consider the $\delta(q^n)$ term appearing in \mref{baxform},
and its meaning in the context of the $T$-$Q$ relations \mref{tqu} and
the commutativity 
with $T$.
For this purpose, it is useful to define a new matrix without 
the delta function: 
\be \tQ(z,\us)\equiv\bQ(z,\us)/\delta(q^n)= 
s_0^{-n/2}  Q_{{Bax}}(v).\lb{tilQ}\ee

For $n=0$, $\delta(q^n)$ is clearly meaningless,
but we conjecture that $\tQ(z,\us)$ still obeys 
the $T-Q$ relations \mref{tqu} and still commutes
with $\bT(z')$: namely  we conjecture that
\begin{equation}\begin{array}{lll}
\bT(z)\,\tQ(z,\us)
&=& \big( \phi_1(z,\us,w ) \big)^N\,
 \tQ(zq^2,\us ^+)\\ &+& \big( \phi _2(z,\us,w)\big)^N \,
\tQ(zq^{-2},\us ^-)  \quad \hbox{and}  \\
\, [ \bT(z') , \tQ(z,\us)] 
&=&0  \lb{tqual}
\end{array}\end{equation}
when $n=0$.  
If Baxter's construction is valid, this should of course be
true, and we have checked up to $N=10$ that the conjecture holds.
Note that for the restriction $s_1=s_2=0$ and the choice 
\mref{rhores} made here, we have
\ben 
\phi_1(z,\us,w)=q\frac{w-z}{w-q^2z},\quad \phi_2(z,\us,w)=1.\een

When $n\neq 0$,  $\delta(q^n)$ is well defined as a formal series in $q$,
and $\bQ(z,\us)$ obeys \mref{tqu} and commutes with $\bT(z')$ by construction. 
As a consequence, it follows that \mref{tqual}
hold whenever $q^n=1$ (checks confirm this up to 
$N=8$). Checks also show that \mref{tqual} are not valid for generic
$q$ and generic $n\neq 0$.

In summary: the situation for $n=0$ is that $\bQ(z,\us)$ is
ill-defined but $\tQ(z,\us)$ appears to obey \mref{tqual}; for
$n\neq 0$, $\bQ(z,\us)$ obeys \mref{tqu} and commutes with $\bT(z')$ by construction, and
$\tQ(z,\us)$ obeys \mref{tqual} when $q^n=1$.

Finally, we show in Appendix B that 
\begin{equation}
[\tQ(z,\us), \tQ(z^\prime,\us^\prime)]=0 \, 
\end{equation}
for all $n$, including $n=0$.
This property implies in particular the commutativity between $\bQ(z,\us)$
and $\bQ(z^\prime, \us ^\prime)$ when $n\neq 0$ (and when we restrict
to $s_1=s_2=0$ as elsewhere in this section).

\newpage
\setcounter{equation}{0}
\section{Discussion}

To summarize: we have defined a $\uqbp$ representation $M(z,\us)$,
 and
used it to construct a $\uqbm$ valued 
operator $\cQ(z,\us)$ 
that obeys the generalized $T$-$Q$ relations
\mref{tq1} and commutes with the operator
$\cT(z')$. In Section 2.3, we have shown how, upon restricting
$\us$, $\cQ(z,\us)$ reduces to the operators $Q_{\pm}(\gl)$
constructed in terms of q-oscillator representations in \cite{AF97}.

We have
then considered a representation of this object on the quantum space 
$V_{w_1}^{(1)} \ot \cdots \ot V_{w_N}^{(1)}$
corresponding to the 6-vertex model, and shown how the $T$-$Q$
relations are modified by the coefficients $\phi_1$ and $\phi_2$
appearing in \mref{tqu}. 

We have gone on to obtain the explicit form \mref{baxform} for
$\bQ(z,\us)$ - valid for the  $s_1=s_2=0$ case. 
When $n\neq 0$, $\bQ(z,\us)$ obeys relations \mref{tqu}, and
commutes with $\bT(z')$ and $\bQ(z',\us')$.
When $n=0$, $\bQ(z,\us)$ is ill-defined, but $\tQ(z,\us)$ obeys
\mref{tqual} (up to $N=10$ at least) and commutes with $\tQ(z',\us)$.
So in either case, $n\neq 0$ or $n= 0$, our construction yields
a well defined $Q$-matrix,
either $\bQ(z,\us)$ or $\tQ(z,\us)$,
that obeys the $T-Q$ relations and commutes both with $\bT(z')$
and with itself at different values of $z$ and $\us$.

In Section 5, we have also discussed the properties of $\tQ(z,\us)$ 
when $n\neq 0$, and there is a  slightly mysterious
aspect to this: while our algebraic construction of $\bQ(z,\us)$ 
is valid for $q$ generic, we find that $\tQ(z,\us)$ obeys relations $\mref{tqual}$
in the root of unity $q^n=1$ case. Clearly, this fact
arises as a consequence of the delta function in \mref{baxform}, that 
in turn  comes from the infinite-dimensional trace. Beyond this
comment, we have no real understanding of this fact, but feel that it 
is still worth pointing it out, especially in the light of recent interest in the 
6-vertex model at roots of unity \cite{FB01}.

We would like to mention some potential
applications of our 
construction that we hope
to consider in future
work. Firstly, there are intriguing 
similarities between our generalized $T-Q$ relations and the functional relations \cite{DT99b}
linking the wave functions of the anharmonic Schr\"odinger equation in different
Stokes sectors. The later functional relations also possess extra parameters
over the conventional $T$-$Q$ relations that shift in a manner very similar 
to ours (although there is one less parameter in the Stokes
relations).
Secondly, it is possible to derive Bethe equations from our 
generalized $T$-$Q$ relations that differ from the conventional 
equations in the 
possible $\us$ dependence of the Bethe roots. It would be
interesting to attempt to understand the solutions of such systems
in the light of the fact that $[\cQ(z,\us),\cQ(z',\us')]\neq 0$ 
in general.
Finally, it would be interesting to consider representations of 
our $\cQ(z,\us)$ on a `continuous' quantum space in order to 
be able to make connections with the constructions given in \cite{BLZb,Pr02}.

\vspace*{5mm}

\nin {\bf Acknowledgements:}
We would like to thank Barry McCoy for his
interest and many useful comments and suggestions.
RAW would like to thank Tetsuji Miwa for
discussions about $\uqbp$ representations. We would also like to
thank Patrick Dorey for pointing out the similarities with 
the functional relations given in \cite{DT99b} and for other
useful observations.
This project was funded  in part by EPSRC grant GR/M97497.

\newpage
\baselineskip=14pt

\baselineskip=17pt
\newpage
\appendix
\setcounter{equation}{0}


\setcounter{equation}{0}
\section{Proof of Proposition 3.1}

Let
$\{a^{(1)}_1,a^{(1)}_2,\cdots,a^{(1)}_N,a^{(2)}_1,a^{(2)}_2,\cdots,a^{(2)}_M\}$
be a basis for $A$,  $\{b_1,b_2,\cdots,b_N\}$ be a basis of $B$, and 
$\{c_1,c_2,\cdots,c_M\}$ be a basis of $C$,
such that the $\cA$ linear injection $\iota$ and surjection $\pi$
are given by
\ben \iota: B \to& A,\quad \quad \pi: &A\to C,\\
            b_i \goto& a^{(1)}_i \quad &a^{(1)}_i \goto 0, \ws 
a^{(2)}_i \goto c_i.\een
The action of $X$ on these basis vectors defines the following
matrices:
\ben
X \,a^{(1)}_j &=& \sli_{i=1}^{N} {^{(1)}}\!X^{A}_{ij} \, a^{(1)}_i,
\quad X \,a^{(2)}_j = \sli_{i=1}^{M} {^{(2)}}\!X^{A}_{ij} \, a^{(2)}_i+
              \sli_{i=1}^{N} {^{(3)}}\!X^{A}_{ij} \, a^{(1)}_i,\\
X \,b_j &=& \sli_{i=1}^{N} X^{B}_{ij} \, b_i,\quad 
X \,c_j = \sli_{i=1}^{M} X^{C}_{ij} \, c_i.
\een
The required trace is given by
\ben \Tr_{A}(X)=\sli_{i=1}^N {^{(1)}}\!X^{A}_{ii} + \sli_{i=1}^M
{^{(2)}}\!X^{A}_{ii}.\een
Now note that we can write
\ben  &&\sli_{i=1}^{N} {^{(1)}}\!X^{A}_{ij} \, a^{(1)}_i = X a^{(1)}_j = X \iota b_j =  \iota X b_j = 
 \sli_{i=1}^{N} X^{B}_{ij} \, a^{(1)}_i, \hb{ and },\\
&& \sli_{i=1}^{M} X^{C}_{ij} \, c_i=  X c^{(1)}_j= X \pi a^{(2)}_j = \pi X a^{(2)}_j =
\sli_{i=1}^{M} {^{(2)}}\!X^{A}_{ij} \, c_i.
\een
Hence, we have 
\ben {^{(1)}}\!X^{A}_{ij}=X^{B}_{ij}, \quad  {^{(2)}}\!X^{A}_{ij}= X^{C}_{ij},
\een
and the proposition then follows.

\newpage

\setcounter{equation}{0}
\section{Proof of $[\tQ(z,\us),\tQ(z^\prime, \us^\prime)]=0$.}

In this appendix we will prove that the operator $\tQ(z, \us)$,
defined in Section 5, has the commutativity property 
\be [\tQ(z,\us), \tQ(z^\prime, \us^\prime)]=0.\lb{qcomr}\ee
(Recall that we have restricted $s_1=s_2=0$ in the definition of $\tQ(z,\us)$.)

From the definition of $\tQ(z,\us)$ given by \mref {Qbax}, \mref
{baxform}, and \mref {tilQ},
it follows that
\be
&&[\tQ(z,\us), \tQ(z^\prime, \us^\prime)]_{\underline {\alpha}}^{\underline {\gamma}}=\nonumber \\
&=&s_0^{-n}\sum _{\underline {\beta}}\left [{\mbox {exp}}\left (\frac {1}{2} i v \, {\underline {\alpha}}\cdot  {\underline {\beta}} 
+\frac {1}{2} i v^\prime \, {\underline {\beta}}\cdot  {\underline {\gamma}}\right )-
{\mbox {exp}}\left (\frac {1}{2} i v^\prime \, {\underline {\alpha}}\cdot  
{\underline {\beta}} 
+\frac {1}{2} i v \, {\underline {\beta}}\cdot  {\underline {\gamma}}\right )\right ]
{\mbox {exp}}\left (\frac {1}{2}i\eta \, ({\underline {\alpha}}-{\underline {\gamma}})\wedge {\underline {\beta}}\right ) \, ,
\nonumber 
\ee
where we have introduced the notation
\begin{equation}
{\underline {\alpha}}\equiv (\alpha _1, \ldots , \alpha _N) \,
\hbox{   etc}, \quad 
{\underline {\alpha}}\cdot  {\underline {\beta}}=\sum _{j=1}^N \alpha _j \beta _j \, , \quad 
{\underline {\alpha}}\wedge {\underline {\beta}}=\sum _{k=1}^N\sum _{j=1}^{k-1}(\alpha _j \beta _k - \alpha _k \beta _j)\, .
\end{equation}
Now, it is simple to show that, given ${\underline {\alpha}}$
belonging to the n-th block (i.e. $\sli_i \ga_i=n$) , every other element ${\underline
  {\gamma}}$ belonging to the same block can be written as
${\underline {\gamma}}=P{\underline {\alpha}}$, where $P$ is an
element of the symmetric group $S^N$ which has the following properties:
\begin{eqnarray}
&&P=P_{i_1,i_2}P_{i_3,i_4}...P_{i_{k-1},i_k} \, , \quad i_1<i_2 \, , \, i_3<i_4 \, , ... \, , \, i_{k-1}<i_k \, , \nonumber \\
&&[i_1,i_2]\supset [i_3,i_4]\supset... \supset [i_{k-1},i_k] \, , \, \quad 1\leq i_1, ... , i_k \leq N \, , \lb{prop21}
\end{eqnarray}
and $P_{j,k}$ is the operator with action 
\ben
P_{j,k}
\, (\alpha_1,\cdots,\alpha_j,\cdots,\alpha_k,\cdots,\ga_N)
=(\alpha_1,\cdots,\alpha_k,\cdots,\alpha_j,\cdots,\ga_N).\een
It follows from the definition of $P$ that the following properties hold
\ben
P^2=1 \, , \quad P {\underline {\alpha}} \cdot P {\underline {\beta}}={\underline {\alpha}} \cdot {\underline {\beta}} \, .
\een
Using these properties we rewrite our commutator in the following form
\be
&&[\tQ(z,\us), \tQ(z^\prime, \us^\prime)]_{\underline {\alpha}}^{P \underline {\alpha}}=\frac {s_0^{-n}}{2}\sum _{\underline {\beta}}\Bigl [{\mbox {exp}}\left (\frac {1}{2} i v \, P {\underline {\alpha}}\cdot P {\underline {\beta}} 
+\frac {1}{2} i v^\prime \, {\underline {\alpha}}\cdot  P{\underline {\beta}}\right )- \nonumber \\
&-&{\mbox {exp}}\left (\frac {1}{2} i v^\prime \, P {\underline {\alpha}}\cdot P {\underline {\beta}} 
+\frac {1}{2} i v \, {\underline {\alpha}}\cdot P {\underline {\beta}}\right )\Bigr ]\left [ {\mbox {exp}}\left (\frac {1}{2}i\eta \, ({\underline {\alpha}}-P{\underline {\alpha}})\wedge {\underline {\beta}}\right )- {\mbox {exp}}\left (\frac {1}{2}i\eta \, ({\underline {\alpha}}-P{\underline {\alpha}})\wedge P {\underline {\beta}}\right ) \right ]\, . \nonumber 
\ee
We will prove by induction (over $P$) the property
\begin{equation}
({\underline {\alpha}}-P {\underline {\alpha}}) \wedge ({\underline {\beta}}-P{\underline {\beta}})=0
\, , \quad \forall\,  {\underline {\alpha}}, \, {\underline {\beta}}, \,  P \, , \lb {Property}
\end{equation}
which implies that the previous commutator is zero. 

Firstly, property \mref {Property} is easily shown when $P=P_{jk}$,
 $1\leq j<k \leq N$. Next, let us make the inductive hypothesis that 
\mref {Property} is true for a given $P$ of the form 
\begin{eqnarray}
&&P=P_{i_1,i_2}P_{i_3,i_4}...P_{i_{k-1},i_k} \, , \quad i_1<i_2 \, , \, i_3<i_4 \, , ... \, , \, i_{k-1}<i_k \, , \nonumber \\
&&[i_1,i_2]\supset [i_3,i_4]\supset... \supset [i_{k-1},i_k] \, , \, \quad 1\leq i_1, ... , i_k \leq N \, , \lb {defP}
\end{eqnarray}
Now define $P'$ by
\be P^\prime =P P_{ij}, \quad\hb{where}\ws i<j \ws\hb{and  }[i_{k-1},i_k] \supset
[i,j]\lb{defPp}.\ee Then we have that
\ben
({\underline {\alpha}}-P^\prime {\underline {\alpha}})=({\underline {\alpha}}-P {\underline {\alpha}})+({\underline {\alpha}}-P_{ij} {\underline {\alpha}})
\een
and so \mref{Property} will be true for $P'$ if
\be
({\underline {\alpha}}-P {\underline {\alpha}}) \wedge ({\underline
  {\beta}}-P_{ij}{\underline {\beta}})+({\underline
  {\alpha}}-P_{ij}{\underline {\alpha}}) \wedge ({\underline
  {\beta}}-P{\underline {\beta}})=0 \, .
\lb{Prop2}\ee
Using the definition of the wedge product, the lhs is equal to
\begin{equation}\begin{array}{lll}
\sum \limits _{i\leq l<j} [\alpha _l - (P\alpha)_l](\beta _j -\beta _i)&-&\sum \limits _{i<l\leq j} [\alpha _l - (P\alpha)_l](\beta _i -\beta _j) - \\
-\sum \limits _{i\leq l<j} [\beta _l - (P\beta)_l](\alpha _j -\alpha _i)&+&\sum \limits _{i<l\leq j} [\beta _l - (P\beta)_l](\alpha _i -\alpha _j)  \, . \lb{lhs21}
\end{array}\end{equation}
However, it follows from the definitions \mref{defP} and \mref{defPp} that we have
\ben
\alpha _l - (P\alpha)_l=0 \,  \quad \forall\, l \ws\hb{such that }
i\leq l\leq j \, , \quad 
\beta _l - (P\beta)_l=0 \,  \quad \forall\, l \ws\hb{such that }
i\leq l\leq j \ \, ,
\een
and therefore that expression \mref{lhs21} is equal to $0$. 
It follows that \mref{Property} and hence \mref{qcomr} is true.

\end{document}